\documentclass[aps,pra,numerical,showpacs,preprint]{revtex4-2}
\usepackage{graphicx}
\usepackage{tikz}
\usetikzlibrary{decorations.pathmorphing,calc}
\usetikzlibrary{arrows.meta}
\usepackage{pgfplots}
\pgfplotsset{compat=1.18} 
\usepackage{subcaption}
\usepackage{caption}
\usepackage[fleqn]{amsmath}
\usepackage{amssymb}
\usepackage{amsfonts}
\usepackage{bbm}
\usepackage{babel}
\usepackage[utf8]{inputenc}
\usepackage{graphicx, float}
\usepackage{rotating}
\usepackage[normalem]{ulem}
\usepackage[hidelinks]{hyperref}
\usepackage{makecell}
\usepackage{diagbox}
\usepackage{array}
\usepackage{dsfont}

\def\bm#1{\hbox{\boldmath$#1$\unboldmath}}

\usepackage{mathtools}
\usepackage{bbold}

\newcommand*\xbar[1]{%
	\hbox{%
		\vbox{%
			\hrule height 0.5pt 
			\kern0.5ex
			\hbox{%
				\kern-0.1em
				\ensuremath{#1}%
				\kern-0.1em
			}%
		}%
	}%
}

\newcommand{\doisp}{\mathrel{\mkern-3mu:\mkern-3mu}}

\newcommand{\muD}{\ensuremath{\mu_{\scriptscriptstyle \Delta}}}
\newcommand{\muS}{\ensuremath{\mu_{\scriptscriptstyle \Sigma}}}
\newcommand{\omD}{\ensuremath{\omega_{\scriptscriptstyle \Delta}}}
\newcommand{\omS}{\ensuremath{\omega_{\scriptscriptstyle \Sigma}}}

\begin{document}

\title{A generalized harmonic oscillator problem for a spin-1/2 fermion}

\author{V. B. Mendrot}
\affiliation{DFI, Physics Department, Unesp - São Paulo State University, Guaratinguetá, Brazil}
\author{A. S. de Castro}
\affiliation{DFI, Physics Department, Unesp - São Paulo State University, Guaratinguetá, Brazil}

\author{P. Alberto}
\affiliation{CFisUC, Physics Department, University of Coimbra,
	P-3004-516 Coimbra, Portugal}

\date{\today}



\begin{abstract}
\noindent

The exact bound-state wavefunctions and the corresponding energy equation are calculated for a new generalized harmonic oscillator problem describing a spin-1/2 fermion in 3+1-dimensions, involving scalar, vector, and tensor couplings acting simultaneously within a particular plane of motion. For the scalar and vector coupling, singular harmonic oscillator shapes are considered, such that the singular term is needed to allow analytical solutions for the wavefunctions while preserving binding under adequate conditions. For the tensor sector,the Dirac oscillator potential is employed, which adds another independent binding mechanism to the problem. The exact bound-state solutions are computed by specifically tuning coefficients for an appropriate pair of \textit{Ansätze} for the radial functions, which leads to wavefunctions in terms of generalized Laguerre polynomials. Although the energy equation cannot provide a general expression for the energy spectrum, specific constraints on the quantum numbers as simple functions of the external potential parameters can be derived for it, which determines the conditions for bound solutions to exist, and of what type: particle, antiparticle or both. It is shown that the results can be simply mapped to the spherically symmetric analogue problem, and this is used to show that the general result encompass several previous particular cases of spherically symmetric harmonic oscillator problems in the Dirac equation available in the literature.
\end{abstract}

\maketitle


\section{Introduction}
\label{Sec:Introduction}

In a previous paper \cite{mendrotGeneralizedCoulombProblem2026}, a new method was developed to decouple radial equations to solve generalized circularly and spherically symmetric Coulomb problems in the 3+1-dimensional Dirac equation which had not yet been achieved. Such method allowed the determination, for the first time in the literature, of the exact characteristics of single spin-1/2 fermion bound states under the simultaneous presence of external scalar, four-vector and tensor Lorentz potentials with Coulomb-like shapes besides the regime of spin or pseudospin symmetries (being that regime obtained as a particular case of the general problem solution). In the present paper,the same method is used to derive results for the case of a generalized harmonic oscillator problem in the Dirac equation which encompasses previous cases available in the literature as particular configurations. Although the methods are mainly the same for both problems, as the reader shall see, the change of the potentials shape requires different types of reasoning during the calculations and subsequent analysis. The results shared here and the ones in the cited reference establish a framework for the examination of simultaneous binding mechanisms originating from scalar, four-vector and tensor potentials for the two main types of potentials known for having analytical solutions in quantum mehcanics.

In the covariant Dirac operator, the potentials can be introduced in several ways due to the different Lorentz structures available \cite{wachterRelativisticQuantumMechanics2011}. The scalar coupling $S(x)$ is introduced by a local mass modification $m+S(x)$, while the four-vector coupling $A_{\mu}(x)=(A_0(x),-\bm{A}(x))$ is introduced by a minimal coupling to the four-momentum operator $p_{\mu}\rightarrow p_{\mu}-A_{\mu}(x)$. Scalar and time-component of vector couplings can be described together in terms of the sum and difference potentials $V_\Sigma=A_0+S$ and $V_\Delta=A_0-S$, which are relevant for discussions regarding spin and pseudospin symmetries (see \cite{ginocchioRelativisticSymmetriesNuclei2005} and references therein). The case of scalar and vector simple harmonic oscillators, for instance, has been considered in the literature, where exact solutions can be found in this regime \cite{Lisboa:2003tp,ginocchioRelativisticHarmonicOscillator2004,deCastro:2006rm}. The remaining components of the four-vector coupling, $\bm{A}$, can be used to describe magnetic fields when the vector coupling is of eletromagnetic origin, and are relevant to describe phenomena such as the Aharonov-Bohm effect, for instance \cite{gitmanSelfAdjointExtensionsQuantum2012}.

Being an exactly solvable model, one of the key features that makes the simple harmonic oscillator potential problem in quantum mechanics so relevant is that it can be used as a good description of low energy dynamics around stable equilibrium points of other more complicated potentials as long as they are sufficiently regular \cite{greinerQuantumMechanicsIntroduction2001}, having therefore wide research applications such as in: molecular vibration \cite{wilsonMolecularVibrationsTheory2012}, atomic traps \cite{buschTwoColdAtoms1998,chenTwoAtomsHarmonic2024,xinRapidQuantumSqueezing2021}, lattice vibration \cite{bornDynamicalTheoryCrystal2023}, particle binding and others. Besides that, its algebraic structure is another relevant fact within the model, since it allows the Hamiltonian to be written in terms of creation and annihilation operators (which are the Hermitean conjugates of each other), being the spectrum quantization entirely determined by their algebra. This is the basis for the construction of normal modes of free quantum fields \cite{peskinIntroductionQuantumField1995}.

An exactly solvable extension of the simple harmonic oscillator problem is the singular harmonic oscillator problem, obtained by the introduction of an inversely quadratic term in $S$ and $A_0$ (or equivalently in their sum and difference potentials), a non-perturbative contribution that imposes a criticality condition under which the bound states are no longer preserved, and are said to \textit{fall to the center} \cite{landauQuantumMechanicsNonrelativistic1991}. Furthermore, questions regarding the preservation of hermiticity of the Hamiltonian arise when one includes this type of potentials \cite{gitmanSelfAdjointExtensionsQuantum2012}.

Besides the scalar and four-vector couplings, a linear tensor potential coupled to the linear momentum operator $\bm{p}\rightarrow\bm{p}-i\beta m\omega\bm{r}$ can also be introduced to obtain an effective harmonic potential with the addition of spin-orbit interaction contributions. This is called the Dirac oscillator \cite{Moshinsky:1989hxi}, which in the non-relativistic limit, recovers the harmonic oscillator spectrum. This coupling has application in nuclear systems \cite{Moshinsky:1989hxi,Alberto:2004kb}. Moreover, there exists  an equivalence between this type of tensor coupling and the spatial component of the four-vector potential ($\bm{A}$) under circular symmetry, as shown in \cite{deCastro:2021ton}. With the Dirac oscillator shape, the azimhutal component of $\bm{A}$ in cylindrical coordinates models a uniform magnetic field transverse to the plane of motion, which is expected, since the Landau levels of a 2+1-dimensional charged particle in a uniform transverse magnetic field can be mapped to the one-dimensional harmonic oscillator \cite{landauQuantumMechanicsNonrelativistic1991}.

Bound-state problems in the Dirac equation involving oscillator-type potentials have been investigated in several different configurations of different degrees of generality in the literature. The first cases developed in the literature were the scalar and time-component of four-vector harmonic osccilators \cite{tegenRelativisticHarmonicOscillator1990} and the pure Dirac oscillator \cite{Moshinsky:1989hxi}. Both cases were subsequently combined, encompassing the spin-symmetric, pseudospin-symmetric, and pure-tensor cases \cite{Lisboa:2003tp}. In this scenario, a Coulomb-like tensor potential was also discussed \cite{akcayDiracEquationScalar2009}, which only shifts the spin-orbit quantum number and does not add any new binding mechanism to the problem. Then, the Cornell-type tensor potential, which is just a combination of Dirac oscillator and Coulomb-like tensor potential was also examined \cite{akcayExactSolutionsDirac2009,zarrinkamarDiracEquationHarmonic2010}, a case mathematically equivalent to the one dealt with in \cite{Lisboa:2003tp}. The singular extension of the scalar-vector oscillator was initially treated without tensor coupling under the spin- and pseudospin-symmetry conditions \cite{aydogduSolutionDiracEquation2009,ikhdairRelativisticNonrelativisticBound2011}. Tensor interactions were subsequently incorporated in the pseudospin-symmetric limit \cite{aydogduExactPseudospinSymmetric2010} and then, more generally, through simultaneous linear and Coulomb-like tensor terms under both spin- and pseudospin-symmetry conditions \cite{hamzaviEXACTLYCOMPLETESOLUTIONS2011}. Finally, a unified class of exact radial Dirac solutions that contains the Dirac oscillator, the ordinary and singular harmonic oscillators, and configurations with linear and Coulomb-like tensor interactions as particular cases was constructed \cite{Garcia:2017xkq}.

All these works share in common the following restrictions: 1) only the time-component of the four-vector potential is non-vanishing; 2) the vector and scalar potentials are such that either $V_\Delta$ or $V_\Sigma$ is constant. The first restriction, as shown in \cite{deCastro:2021ton}, can be directly incorporated in the known solutions when in presence of the Dirac oscilator potential, both in circularly or spherically symmetric scenarios by simply reintrepeting the potential parameters to include the contribution of the vector potential. As for the latter setup, those are exactly the spin and pseudospin symmetry conditions, respectively. Therefore, in all previously considered scenarios, the unfolding of analytical solutions is constrained by the imposition of spin or pseudospin symmetry.

The consideration of simultaneous scalar and vector simple harmonic oscillator potentials outside the conditions of spin or pseudospin symmetry is known to not allow exact solutions due to the presence of a fourth power term in the effective potential arising from the cross-term $V_\Delta V_\Sigma$ when decoupling the equations, which only vanishes under spin or pseudospin symmetry \cite{Lisboa:2003tp}. However, if one introduces a singular term in the potentials, there can be two possible configurations under which the fourth power term (and other problematic terms) vanishes beyond the condition of spin or pseudospin symmetries: singular sum potential and simple harmonic oscillator difference potential; simple harmonic oscillator sum potential and singular difference potential. Under these two configurations, both scalar and vector potentials are single harmonic potentials, there being no need to impose spin or pseudospin symmetries conditions upon them. In this work, it is shown that these two configurations (which are connected by a charge conjugation transformation) plus the Cornell tensor potential (Dirac oscillator plus Coulomb-like shape) can be solved exactly in terms of generalized Laguerre polynomials for the radial functions, and an exact irrational energy equation can be obtained. This problem, which is qualitatively distinct from the ones covered in the literature, enlarges the possible descriptions of harmonic oscillator problems in the Dirac equation, and constitutes another possible generalization of some of the previous cases, which are discussed further in the text. Besides being a new generalization, its new feature is that it does not rely on the need to impose spin or pseudospin symmetry conditions.

The geometric setup considered is that the potentials act only on a specific plane of motion. The 3+1-dimensional Dirac equation is solved with the condition that $p_z\Psi=0$ \cite{deCastro:2021ton}. This is suitable for scenarios where the motion is restricted to the xy-plane. The main interest in this work is devoted to potentials which are circularly symmetric, i.e., they depend only on the radial coordinate in the plane of motion. Therefore the Hamiltonian is written in cylindrical coordinates. In this work, the same formalism of Ref. \cite{deCastro:2021ton} is used, which is better suited to the main goal of this work because it maintains a close relationship with the formalism for spherically symmetric Dirac spinors and thus facilitates the comparison between the features of circularly symmetric Hamiltonians and their eigenspinors with the spherically symmetric ones, which is also discussed here. A systematic way of identifying viable bound states from the solutions found is proveided by analyzing the conditions for the potential parameters and quantum numbers to yield binding from the energy equation obtained.

This work is organized as follows: in section \ref{sec:Dirac}, the notation and mathematical setup for the problem is built. Section \ref{sec:osciladorgeral} is the main part of the work, in which the potentials are presented, the \textit{Ansätze} for the radial functions are built, and the equation for the radial functions is solved. Then, the quantum condition originated from the radial equation is analyzed to determine the conditions on the spectrum and the classification of the bound-states. In \ref{sec:particular}, the particular cases of the problem are recovered from the generalized problem solutions and the correponding spectrum analysis is obtained from the general analysis. Finally, in section \ref{sec:conclusion}, the conclusions from the work are drawn.

	\section{The general setup for a circularly symmetric problem in the Dirac equation}
	\label{sec:Dirac}
	
	Under the simultaneous presence of time-independent scalar $S$, four-vector $A^{\mu}=(A_0,\bm{A})$ and Dirac oscillator-type tensor $\bm{U}$ potentials, the  Dirac hamiltonian in a 3+1-dimensional Minkowski space-time for a single mass $m$ spin-1/2 fermion is given by ($\hbar=c=1$)
	
	\begin{gather}\label{Ham}
		H=\bm{\alpha}\cdot(\bm{p} - \bm{A}) + i\beta\bm{\alpha}\cdot\bm{U}  + \beta(m + S) + V.
	\end{gather}
	
	\noindent The Dirac matrices $\alpha_i$ and $\beta$ are 4$\times$4 and obey the algebra $\left\lbrace \alpha_i,\alpha_j\right\rbrace=2\delta_{ij}\mathbb{1}$, $\left\lbrace \alpha_i,\beta\right\rbrace=0$, $\alpha_i^2=\beta^2=\mathbb{1}$. In their standard representation, they are
	
	\begin{gather}
		\alpha_i = \begin{pmatrix}
			0 & \sigma_i \\
			\sigma_i & 0
		\end{pmatrix}
		\quad,\quad
		\beta = \begin{pmatrix}
			\mathbb{1}_2 & 0 \\
			0 & -\mathbb{1}_2
		\end{pmatrix}
	\end{gather}
	
	\noindent in which $\sigma_i$, $i=1,2,3$ are the Pauli matrices.
	
	To determine the solutions of the Dirac equation, one must solve the eigenvalue equation $H\Psi(\bm{x})=\varepsilon\Psi(\bm{x})$, where $\Psi(\bm{x})$ is the physical system wavefunction and $\varepsilon$ is its total energy. For bound-state solutions, $\int d^3 \bm{x} \Psi^{\dagger}(\bm{x})\Psi(\bm{x})$ must be finite, such that it can be chosen to be normalized to unity. The investigation intended is of solutions for when all the potentials abide circular symmetry. For that, it is considered that $p_z\Psi=0$, as in Ref. \cite{deCastro:2021ton}. The potentials are taken to be central and such that the spin-orbit operator $K=\beta\left( L_z\Sigma_z + 1/2\right)$ commutes with the Hamiltonian (see \cite{mendrotSymmetryGeneratorsQuantum2025}). Choosing to work with the cylindrical coordinates system $(\rho,\phi,z)$, the conditions upon the potentials become $A_0=A_0(\rho)$, $\bm{A}=A_\rho(\rho) \bm{\hat{\rho}} + A_\phi (\rho)\bm{\hat{\phi}}$, $S=S(\rho)$ and $\bm{U}=U_\rho(\rho)\bm{\hat{\rho}}$. This is important to obtain the radial functions for the equivalent problem with spherical symmetry by just a simple map, as shown later in the text. The potential $A_\rho$ does not introduce any independent structure to the problem, since it appears in the Hamiltonian as part of a radial covariant derivative, and therefore can be gauged away by a redefinition of a local phase in the spinor, as will be done in (\ref{spinor}), simplifying the equations without altering any observable feature of the system. Following Ref. \cite{deCastro:2021ton}, the spinor is written as
	
	\begin{gather}\label{spinor}
		\Psi_{km_j}=\dfrac{1}{\sqrt{\rho}}\left(\begin{matrix}
			ig_{k}(\rho)h_{km_j}(\phi)\\[5pt]
			f_{k}(\rho)h_{-km_j}(\phi)
		\end{matrix}\right) \textit{exp}\left(-i\int^{\rho}A_{\rho}(\rho')d\rho'\right) 
		\quad,\quad
	\end{gather}
	
	\noindent which is a simultaneous eigenstate of the spin-orbit operator $K=\beta\left( L_z\Sigma_z + 1/2\right)$ and the z-component of the total angular momentum operator $J_z=L_z+\Sigma_z/2$. Both their eigenvalues can only take semi-integer values: $k,m_j=\pm 1/2,\pm 3/2,\pm 5/2,...$, such that $k=\pm m_j$, due to the fact that $K=\mathcal{S}_zJ_z$, where $\mathcal{S}_z=\beta\Sigma_z$ is another mutually commuting operator with eigenvalues $\pm 1$ \cite{mendrotSymmetryGeneratorsQuantum2025}. The angular part of the spinor is contained in the spinorial circular harmonics $h_{km_j}=\Phi_l\chi_s$, constructed by composing the $L_z$ eigenstates $\Phi_l=e^{il\phi}/\sqrt{2\pi}$ with the eigenstates of the $z$-component of the 2$\times$2 spin operator $\sigma_z$, $\chi_s=\left(\begin{matrix} \delta_{s,1} & \delta_{s,-1}\end{matrix}\right)^T$, $s=\pm 1$. These harmonics have two relevant properties: $\int_{0}^{2\pi}d\phi\; h^{\dagger}_{k^{'}m^{'}_j}h_{km_j}=\delta_{k^{'}k}\delta_{m^{'}_jm_j}$ and $\bm{\sigma}\cdot\hat{\bm{\rho}} h_{km_j}=h_{-km_j}$. The eigenvalues are related by $k=m_js$ and $m_j=l+s/2$.

	Writing $V_\Sigma=A_0+S$ and $V_\Delta=A_0-S$, the radial equations are
	
	\begin{gather}
		\dfrac{dg}{d\rho} - \dfrac{k}{\rho}g + \tilde{U}g = (m + \varepsilon - V_\Delta)f,\label{g1}\\[5pt]
		\dfrac{df}{d\rho} + \dfrac{k}{\rho}f - \tilde{U}f = (m - \varepsilon + V_\Sigma)g,\label{f1}
	\end{gather}
	
	\noindent where $\tilde{U}=U_{\rho} + (k/m_j)A_\phi$. Since $A_\phi$ and $U_\rho$ act in the same way \cite{deCastro:2021ton}, from now on, their unified aspect of a ``tensor-vector'' potential $\tilde{U}$ is considered. Equations (\ref{g1}) and (\ref{f1}) are related via charge conjugation, which yields the following transformations: $f \leftrightarrow g$, $k \leftrightarrow -k$, $m_j\leftrightarrow m_j$, $\varepsilon \leftrightarrow -\varepsilon$, $\tilde{U} \leftrightarrow -\tilde{U}$ and $V_\Delta \leftrightarrow -V_\Sigma$.
	
	\subsection{Mapping to the spherically symmetric analog problem}
	
		In the spherical symmetry setup, the spinor is
	
	\begin{gather}
		\Psi_{{k_s}m}=\dfrac{1}{r}\left(\begin{matrix}
			ig_{k_s}(r)\Omega_{{k_s}{m_j}}(\bm{\hat{r}})\\[5pt]
			-f_{k_s}(r)\Omega_{{-k_s}{m_j}}(\bm{\hat{r}})
		\end{matrix}\right)
		\quad,\quad
	\end{gather}
	
	\noindent in which $\Omega_{{k_s}{m_j}}(\bm{\hat{r}})$ are the spinor spherical harmonics \cite{greinerRelativisticQuantumMechanics2013a}. The spinor is built to be an eigenvector of the spherical spin-orbit operator $K_s=\beta\left(\bm{L}\cdot\bm{\Sigma} + 1\right)$ and $J_z$ with eigenvalues $k_s=\pm1,\pm2,...$ and $m_j=\pm1/2,\pm3/2,...$, respectively. The radial equations are \cite{Garcia:2017xkq}

	\begin{gather}
		\dfrac{dg}{dr} + \dfrac{k_s}{r}g + \tilde{U}g = (m + \varepsilon - V_\Delta)f,\\[5pt]
		\dfrac{df}{dr} - \dfrac{k_s}{r}f - \tilde{U}f = (m - \varepsilon + V_\Sigma)g.
	\end{gather}
	
	\noindent Comparison to the original spinor (\ref{spinor}) and radial equations (\ref{g1}) and (\ref{f1}) shows that the problem solved in this paper can be directly mapped to the spherical counterpart by taking $k\rightarrow -k_s$ and $h_{km_j}\rightarrow\Omega_{{k_s}{m_j}}$ in the solutions.
	
	\section{The general harmonic oscillator problem}
	\label{sec:osciladorgeral}
	
	Considering the potentials $S$ and $A_0$ to have the shape of circularly symmetric singular harmonic oscillator potentials
	
	\begin{gather}
		S(\rho)=\frac{\mu_{\scriptscriptstyle S}}{\rho^2}+\omega_{\scriptscriptstyle S}\rho^2 \quad,\quad A_0(\rho)=\frac{\mu_{\scriptscriptstyle V}}{\rho^2}+\omega_{\scriptscriptstyle V}\rho^2,
	\end{gather}
	
	\noindent their sum and difference potentials will have this shape as well:
	
	\begin{gather}
		V_\Sigma(\rho)=A_0(\rho)+S(\rho)=\frac{\muS}{\rho^2}+ \omS \rho^2 \quad,\quad V_\Delta(\rho)=A_0(\rho)-S(\rho)=\frac{\muD}{\rho^2}+\omD\rho^2,
	\end{gather}
	
	\noindent in which $\mu_{\scriptscriptstyle \Sigma/\Delta}=\mu_{\scriptscriptstyle V}\pm\mu_{\scriptscriptstyle S}$ and $\omega_{\scriptscriptstyle \Sigma/\Delta}=\omega_{\scriptscriptstyle V}\pm\omega_{\scriptscriptstyle S}$.
	
	As discussed in the introduction, this combination of potentials will, in general, lead to problems in determining analytical bound-solutions, due to the appearence of terms proportional to $\rho^{\pm 4}$ in the second order equations for the radial functions, due to the cross term $V_\Delta V_\Sigma$ (see equations (64) and (65) of Ref. \cite{mendrotSymmetryGeneratorsQuantum2025}). This can be overcome by considering spin or pseudospin symmetries conditions, when either $V_\Delta$ or $V_\Sigma$ is constant, respectively. However, there are two other configurations which have not yet been reported that lead to exactly solvable equations: when I) $\muS=0$, $\omD=0$; and II) $\muD=0$, $\omS=0$. Under these two configurations, the original scalar and vector potentials remain having the singular harmonic oscillator shape while the spin and pseudospin symmetries conditions are violated. Therefore, such configuration is qualitative different from the problems solved in the literature, as discussed in the Introduction. In this work, the solutions to this setup are obtained and analyzed to discuss the new features that arise in this generalized problem. Furthermore, it is proven that they contain previous problem solutions as particular cases, and the conditions the parameters of these problems must satisfy are determined by particularizing the results of the generalized case analysis, therefore filling the gaps in the literature for each one of these problems.
	
	 Since configurations I and II can be related through a charge conjugation transformation of the equations, only the first one will be developed here, which also furnishes the solution to the other case by taking the charge conjugation transformation described after equation (\ref{f1}). Now, considering the tensor-vector sector as well, the potentials of the problem are
	
			\begin{gather}
		V_\Sigma(\rho)=\omS \rho^2 
		\;,\;
		V_\Delta(\rho)=\frac{\muD }{\rho^2}
		\;,\;
		\tilde{U}(\rho)=\frac{a}{\rho} + b\rho.
	\end{gather}
	
	\noindent The tensor-vector sector is considered with a Cornell-type potential shape. Besides it being the most general shape found in the literature for harmonic oscillator problems, it must be emphasized that the solution presented here also considers its four-vector origin, which, for example, models a constant magnetic field transverse to the plane of motion if $A_\phi$ is to be an eletromagnetic potential.
	
	The radial equations (\ref{g1}) and (\ref{f1}) become
	
	\begin{gather}	
		\dfrac{dg}{d\rho}=-b\rho g + \dfrac{\bar{k}}{\rho}g + \left(m + \varepsilon - \dfrac{\muD }{\rho^2}\right)f, \label{dg}
		\\[5pt]
		\dfrac{df}{d\rho}=b\rho f - \dfrac{\bar{k}}{\rho}f + \left(m - \varepsilon + \omS  \rho^2 \right)g, \label{df}
	\end{gather}
	
	\noindent in which $\bar{k}=k-a$. In this configuration, both tensor-vector sector (Dirac oscillator plus constant magnetic field) and the sum harmonic potential have binding possibilities, which are affected by anti-binding behaviour of the singular difference potential. After solving the eingenvalue equation for the Hamiltonian, the exact conditions for the binding to happen, and wether it binds particles, antiparticles or both types of state will be shown.
	
	\subsection{Radial functions}
	
	In the asymptotic limit as the radial functions $g$ and $f$ go to infinity, one sees from equations (\ref{dg}) and (\ref{df}) that they must have the exponential forms $g=Ae^{\nu\rho^2}$ and $f=Be^{\lambda\rho^2}$, where $\nu$ and $\lambda$ are real negative numbers. If the exponents are to be the same, then $\nu=\lambda=-\sqrt{b^2+\omS (m+\varepsilon)}/2$, such that $b^2+\omS (m+\varepsilon)>0$. If the exponents are different, the equation they must satisfy is
	$\nu\lambda - (\nu-\lambda)b=b^2+\omS (m+\varepsilon)$, which can be shown to be, besides one exception, violated if the right-hand side is less than or equal to zero. The exceptions occur when $\varepsilon=-m$ for $b<0$, in which $g=0$ and $f\propto e^{b\rho^2/2}$, and when $\varepsilon=m$ for $b>0$, in which $f=0$ and $g\propto e^{-b\rho^2/2}$. However, as will be shown later, these cases are also included in the \textit{Ansätze} with equal exponents for both radial functions. Therefore, it can be assumed without loss of generality that $b^2+\omS (m+\varepsilon)>0$ and the \textit{Ansätze} with equal exponents can be used thoroughly. This motivates the change of variables

	\begin{gather}
		y=\chi^2\rho^2
	\quad,\quad
	\chi=\left[b^2+\omS (1+E)\right]^{1/4},
	\end{gather}
	
	\noindent such that $b^2+\omS (1+E)>0$. Hereafter, the potentials parameters are now scaled to be dimensionless:
	
	\begin{gather}
		\frac{\varepsilon}{m}=E \quad,\quad\frac{b}{m^2}\rightarrow b \quad,\quad \frac{\omS }{m^3}\rightarrow\omS  \quad,\quad m\muD \rightarrow\muD ,
	\end{gather} 
	
	 The asymptotic behavior towards infinity becomes $g=Ae^{-y/2}$ and $f=Be^{-y/2}$. The new equations are then
	
		\begin{gather}
		2\dfrac{dg}{dy}=\left(\frac{\bar{k}}{y} - \frac{a}{\chi^2}\right)g + \left(\frac{1+E}{\chi\sqrt{y}} - \frac{\muD \chi}{y^{3/2}}\right)f ,  \label{g}
		\\[5pt]
		2\dfrac{df}{dy}=\left(\frac{\omS  \sqrt{y}}{\chi^3} + \frac{1-E}{\chi\sqrt{y}}\right)g - \left(\frac{\bar{k}}{y} - \frac{a}{\chi^2}\right)f. \label{f}
	\end{gather}
	
	Moving on to the analysis of the behaviour of the radial functions near the origin of the coordinates, an uncoupled second-order equation for $f$ can be obtained, which is solved by $f=Ay^{(\delta+1)/4}+By^{-(\delta-1)/4}$, in which $\delta=2\sqrt{(\bar{k}+1/2)^2-\muD  (1-E)}$. By considering this behaviour in the energy decomposition equations (\ref{exp+}) and (\ref{exp-}) given in Appendix \ref{ap:decomp}, the only solution that can mantain the expected values finite and real is $f=Ay^{(\delta+1)/4}$, such that $\delta>2$. Then, $g=By^{(\delta-1)/4}$, and $(\bar{k}+1/2)^2-\muD  (1-E)>1$.
	
	Thus, let the the general pair of \textit{Ansätze} be
	
	\begin{gather}
		g(y)=\alpha\;y^{(\delta-1)/4}e^{-y/2}\left[G(y)+F(y)\right], \label{ansatzg}\\[5pt]
		f(y)=\beta\;y^{(\delta+1)/4}e^{-y/2}\left[G(y)-F(y)\right], \label{ansatzf}
	\end{gather}
	
	\noindent in which $\alpha$ and $\beta$ are undetermined non-zero constants, and $F$ and $G$ are functions of $\tilde{\rho}$. Since $F$ and $G$ are linear combinations of the original radial functions, no change has been made regarding the completeness of the solutions. To preserve the conditions near the origin for $g$ and $f$ , $F$ and $G$ must be constants in the limit as $\tilde{\rho}\rightarrow 0$. For the asymptotic behavior towards infinity, $F$ and $G$ must obey, at most, $\lim_{\tilde{\rho}\rightarrow \infty}F,G \rightarrow \exp(\Lambda\tilde{\rho}^{\epsilon})$, $\epsilon<1$.
	
	The reader should note that neither the constants $\alpha$ and $\beta$ are the constant obtained from the linearity of the solutions of the original equation, which is given by $C_{nk}$ in (\ref{FLag}) below. Furthermore, it was chosen for them to have non-zero values by construction, since the solutions in which any of the radial functions are zero will already be contained in this \textit{Ansätze}, as will be shown later.
	
	Substituting (\ref{ansatzg}) and (\ref{ansatzf}) in equations (\ref{g}) and (\ref{f}), new equations for $G(y)+F(y)$ and $G(y)-F(y)$ are obtained. Adding and subtrating those equations yields
	
	\begin{gather}
		\frac{dG(y)}{dy} - \left(\frac{1}{2} + \Lambda^+ - \frac{\delta/4-\Pi^-}{y}\right) G(y) +\left(\frac{b}{2\chi^2} - \Lambda^- - \frac{\bar{k}+1/2 + 2\Pi^+}{2y}\right) F(y)=0, \label{dG}
	\\[5pt]
		\frac{dF(y)}{dy} - \left( \frac{1}{2} - \Lambda^+ - \frac{\delta/4+\Pi^-}{y}\right) F(y) + \left( \frac{b}{2\chi^2} + \Lambda^- - \frac{\bar{k}+1/2 - 2\Pi^+}{2y}\right)  G(y)=0, \label{dF}
	\end{gather}
	
	\noindent in which the following quantities arise:
	
	\begin{gather}
		\Lambda^{\pm}=\frac{1}{4}\frac{\alpha}{\beta}\frac{\omS }{\chi^3} \pm \frac{1}{4}\frac{\beta}{\alpha}\frac{1+E}{\chi},
	\\[5pt]
		\Pi^{\pm}=\frac{1}{4}\frac{\alpha}{\beta}\frac{1-E}{\chi}\pm\frac{1}{4}\frac{\beta}{\alpha}\muD  \chi.
	\end{gather}
	
	A second order coupled equation for $F(y)$ can be obtained by differentiating (\ref{dF}) and eliminating the first derivative in $G(y)$ via substitution with equation (\ref{dG})
	
	\begin{gather}
	\begin{aligned}
		&\frac{d^2F(y)}{dy^2} - \left( \frac{1}{2} - \Lambda^+ - \frac{\delta/4+\Pi^-}{y}\right) \frac{dF(y)}{dy}
	\\[2pt]
		& \hspace{-0.5cm}-\left[\frac{\delta/4+\Pi^-}{y^2} + \left( \frac{b}{2\chi^2} + \Lambda^- - \frac{\bar{k}+1/2 - 2\Pi^+}{2y}\right) \left( \frac{b}{2\chi^2} - \Lambda^- - \frac{\bar{k}+1/2 + 2\Pi^+}{2y}\right) \right] F(y)
	\\[2pt]
		& \hspace{-0.5cm}-\left[\frac{\bar{k}+1/2 - 2\Pi^+}{2y} - \left(\frac{1}{2} + \Lambda^+ - \frac{\delta/4-\Pi^-}{y}\right) \left( \frac{b}{2\chi^2} + \Lambda^- - \frac{\bar{k}+1/2 - 2\Pi^+}{2y}\right) \right] G(y)=0.
	\end{aligned}\label{G2}
	\end{gather} 
	
	\noindent The last term of the coupling term can be substituted with (\ref{dF}) to eliminate its dependence on $G(y)$. The remaining coupling term is a function of the ratio $\alpha/\beta$ and therefore gives the uncoupling condition
	
	\begin{gather}
		2\Pi^+=\bar{k}+1/2
	\end{gather}
	
	\noindent Solving for the ratio, one finds:
	
	\begin{gather}\label{unccond}
		\frac{\alpha}{\beta}=\frac{(2\bar{k}+1 \pm \delta)\chi}{2(1-E)}.
	\end{gather}
	
	\noindent There are two different and equally valid possibilities of uncoupling. The one with the negative sign is chosen, since it leads directly to a Kummer equation for $F(y)$, of which the solution is easily recognizable. Therefore, equation (\ref{G2}) can be written as
	
	\begin{gather}
		y\frac{d^2F(y)}{dy^2}+\left(\frac{\delta}{2} - y\right)\frac{dF(y)}{dy} - \left[ \frac{\delta}{4} + \frac{1 - E^2 - (2k+1)a - \omS \muD }{4\chi^{2}}\right] F(y)=0.
	\end{gather}
	
	The quantization condition that leads to regular and normalizable solutions for $F(y)$ can now be identified, and is given by
	
	\begin{gather}
		\frac{\delta}{4} + \frac{1 - E^2 - (2k+1)a - \omS \muD }{4\chi^{2}}=-n \;,\; n=0,1,2,3...
		\label{energyeq}
	\end{gather}
	
	\noindent and finally, the solutions for the radial function will be given in terms of generalized Laguerre polynomials
	
	\begin{gather}
		F_{n}(y)=C_{nk}L^{(-1+\delta/2)}_{n}(y).  \label{FLag}
	\end{gather}
	
	\noindent From equation (\ref{dF}), $G(y)$ is also obtained:
	
	\begin{gather}
		G_{n}(y)=C_{nk}\left[\frac{8\sigma\chi^2(1-E)}{\Delta(\sigma)}L^{(\delta/2)}_{n-1}(y) - \frac{\Omega(\sigma)}{\Delta(\sigma)}L^{(-1+\delta/2)}_{n}(y)\right],
	\end{gather}
	
	\noindent such that
	
	\begin{gather}
		\sigma=2\bar{k}+1 - \delta,
	\\[5pt]
		\Delta(\sigma)=\omS \sigma^2+4b(1-E)\sigma - 4(1-E)^2(1+E),
	\\[5pt]
		\Omega(\sigma)=	\omS \sigma^2-4\chi^2(1-E)\sigma + 4(1-E)^2(1+E),
	\end{gather}
	
	\noindent subject to the restrictions $\sigma\neq0$ and $\Delta(\sigma)\neq 0$. Finally, the complete radial functions (\ref{ansatzg}) and (\ref{ansatzf}) are
	
	\begin{gather}
		g(y)=\bar{\alpha}\;y^{\frac{\delta-1}{4}}e^{-y/2}\left[\frac{\Delta(\sigma)-\Omega(\sigma)}{\Delta(\sigma)}L^{(-1+\delta/2)}_{n}(y) + \frac{8\sigma\chi^2(1-E)}{\Delta(\sigma)}L^{(\delta/2)}_{n-1}(y) \right],
	\\[5pt]
		f(y)=\bar{\beta}\;y^{\frac{\delta+1}{4}}e^{-y/2}\left[\frac{\Delta(\sigma)+\Omega(\sigma)}{\Delta(\sigma)}L^{(-1+\delta/2)}_{n}(y) - \frac{8\sigma\chi^2(1-E)}{\Delta(\sigma)}L^{(\delta/2)}_{n-1}(y) \right] ,
	\end{gather}
	
	\noindent with $\bar{\alpha}=C_{nfk}\alpha$ and $\bar{\beta}=-C_{nk}\beta$. Bound solutions must not violate $\sigma\neq0$ and $\Delta(\sigma)\neq0$.
	
	\subsection{Eigenergy spectrum}
	
	The energy equation (\ref{energyeq}) yielded by the quantization condition of the problem must be further analyzed. Defining the energy equation function:
	
	\begin{gather}
		S(E)=E^2-\Gamma -\left[4n+2\sqrt{B(E)}\right] \sqrt{A(E)},\label{eqenergy}
	\end{gather}
	
	\noindent in which
	
	\begin{gather}
		A(E)=b^2+\omS (E+1), \label{A}
		\\[5pt]
		B(E)=\tilde{k}^2+\muD (E-1) \quad,\quad \tilde{k}=\bar{k}+1/2, \label{B}
		\\[5pt]
		\Gamma=1-\omS \muD -2\tilde{k}b,
	\end{gather}
	
	\noindent the eigenergies are then the real roots of $S(E)$. A general expression for them cannot be found, since the algebraic equation obtained from the energy equation is an eight-degree equation in $E$. However, one can extract information regarding the spectrum behavior without solving it by analyzing the roots of $S(E)$. It is shown below that there are at most two real roots, such that the conditions for a given number of them to appear can be cast very simply in form of inequalities involving the quantum numbers and simple functions of the potentials parameters. Therefore, a complete reference guide to the character of the bound states for any configuration of the potentials parameters is achieved.
	
	\subsubsection{Domain analysis, root counting and classification of states}
	
	The conditions upon the parameters of the problem and the quantum numbers are encoded in the restrictions: $A(E)>0$, $B(E)>1$ and $\Gamma<E^2$. The domain of $S(E)$ which corresponds to the problem is then $\mathcal{D}(S(E))\doisp\mathcal{H} \cap \left\lbrace E | E^2>\Gamma\right\rbrace$, in which a connected subdomain, $\mathcal{H}\doisp\left\lbrace E | A(E)>0\right\rbrace \cap \left\lbrace E | B(E)>1\right\rbrace$, intersect with another subdomain that can be either connected or disconnected, depending on the sign of $\Gamma$. Therefore, the domain can be of three types: two-branched, being it two disconnected semi-lines or line segments; one-branched, being it the real line, a semi-line or a line segment; and the empty set, which is interpreted as there being no possible bound-states.

	The connected subdomain $\mathcal{H}$ has as its left-most value $L=\max\left\lbrace L_A,L_B\right\rbrace$, and as its right-most value $R=\min\left\lbrace R_A,R_B\right\rbrace$, where:
	
	\begin{gather}
		\label{LRA}
		\hspace{-10pt}L_A=\left\lbrace
		\begin{alignedat}{2}
			 -&1-\dfrac{b^2}{\omS } \;&&,\;\text{if }\omS >0
			\\[5pt]
			& -\infty \;&&,\;\text{if }\omS \leq0
		\end{alignedat}
		\right.
		\quad,\quad
		R_A=\left\lbrace
		\begin{alignedat}{2}
			& \infty \;&&,\;\text{if }\omS \geq0
			\\[5pt]
			 -1&-\dfrac{b^2}{\omS } \;&&,\;\text{if }\omS <0
		\end{alignedat}
		\right.,
	\end{gather}
	
	\noindent such that when $\omS =0$, $b$ cannot be zero; and
	
	\begin{gather}
		\label{LRB}
		\hspace{-10pt}L_B=\left\lbrace
		\begin{alignedat}{2}
			1&+\dfrac{1-\tilde{k}^2}{\muD } \;&&,\;\text{if }\muD >0
			\\[5pt]
			& -\infty \;&&,\;\text{if }\muD \leq0
		\end{alignedat}
		\right.
		\quad,\quad
		R_B=\left\lbrace
		\begin{alignedat}{2}
			& \infty \;&&,\;\text{if }\muD \geq0
			\\[5pt]
			1+&\dfrac{1-\tilde{k}^2}{\muD } \;&&,\;\text{if }\muD <0
		\end{alignedat}
		\right..
	\end{gather}
	
	\noindent with the additional condition that when $\muD =0$, $\left|\tilde{k}\right|>1$. The sector $\mathcal{H}$ is then
	
	\begin{gather}
		\mathcal{H}:\left\lbrace\begin{alignedat}{2}
			(L &, R) \;&&,\; L < R
			\\[5pt]
			&\emptyset \;&&,\; L \geq R
		\end{alignedat} \right. .
		\label{H}
	\end{gather}

	Inside $\mathcal{H}$, 
	
	\begin{gather}
		S''(E)=2 + \frac{n\,\omS ^{2}}{A(E)^{3/2}} + \frac{1}{2}\,\frac{\left[\muD  A(E) - \omS  B(E)\right]^{2}}{A(E)^{3/2} B(E)^{3/2}}>0,
	\end{gather}
	
	\noindent therefore $S'(E)$ is a monotonically increasing function and $S(E)$ is a convex function inside it. Thus, $S(E)$ can have at most two roots (eigenenergies), which is physically expected, since there can be at most a one-particle and an one-antiparticle solution for a given configuration of parameters. 
	
	Regarding the intersection of $\mathcal{H}$ with $\left\lbrace E | E^2>\Gamma\right\rbrace$, which is the domain of the problem: if $\Gamma<0$, then the domain $\mathcal{D}$ coincides with $\mathcal{H}$; whenever $\Gamma\geq0$, the second set introduces a $2\sqrt{\Gamma}$ wide gap around $E=0$ which possibly intersects $\mathcal{H}$. However, in these cases the gap is entirely in the negative semi-plane of $S(E)$, since $S(\pm\sqrt{\Gamma})<0$. Therefore, it never contains any root of $S(E)$. Thus, an important conclusion is drawn: the number of roots of $S(E)$ can be counted by only considering its behaviour inside $\mathcal{H}$. In other words, the restriction $\Gamma<E^2$ is actually irrelevant for the analysis of the roots. It is relevant, however, in the determination of the conditions for $\mathcal{D}$ to be empty, which is another condition for there not to be bound states, besides there being no real roots in a non-empty domain.
	
	In Appendix \ref{ap:S(E)}, it is proven an important general feature of $S(E)$: whenever it admits a stationary minimum inside $\mathcal{H}$, it is necessarily non-positive. Furthermore, when it is exactly zero, the only root has multiplicity two. This means that it is a case where both particle and antiparticle states are admissible, such that both have the same energy. Knowing these facts, and considering that $S(E)$ is a convex function inside $\mathcal{H}$, one can completely distinguish the scenarios under which there will be a certain number of roots by evaluating the sign of $S(E)$ and $S'(E)$ in the boundaries of $\mathcal{H}$. In Appendix \ref{ap:S(E)}, it is shown in details which combination of signs leads to each number of solutions. From these combination of signs, conditions on the quantum numbers $n$ and $\tilde{k}$ can be extracted in terms of the external potentials parameters.
	
	It is convenient to investigate the conditions yielded by the combinations of signs in Table \ref{TabRoots} for each combination of signs of $\omS$ and $\muD$. From (\ref{LRA}) and (\ref{LRB}), one finds
	
	\begin{table}[ht]
		\centering
		
		\setcellgapes{10pt}
		\makegapedcells
		
		\begin{tabular}{c|c|c}
			& $\muD>0$
			& $\muD<0$ \\
			\hline
			
			$\omS>0$
			&
			$\;
			\left( \max\left\{
			-1-\dfrac{b^2}{\omS},
			\,
			1+\dfrac{1-\tilde{k}^{\,2}}{\muD}
			\right\},
			\infty\right) 
			\;$
			&
			$
			\left( -1-\dfrac{b^2}{\omS},
			\,
			1+\dfrac{1-\tilde{k}^{\,2}}{\muD}\right) 
			$
			\\
			\hline
			
			$\omS<0$
			&
			$
			\left( 1+\dfrac{1-\tilde{k}^{\,2}}{\muD},
			\,
			-1-\dfrac{b^2}{\omS}\right) 
			$
			&
			$\;
			\left( -\infty,
			\,
			\min\left\{
			-1-\dfrac{b^2}{\omS},
			\,
			1+\dfrac{1-\tilde{k}^{\,2}}{\muD}
			\right\}\right) 
			\;$
			\\
			\hline
		\end{tabular}
		
		\caption{Subdomain $\mathcal{H}$ according to the signs of $\omS$ and $\muD$.}
		\label{tab:domain-limits}
	\end{table}
	
	\noindent An analysis of all the combinations of Table \ref{TabRoots} for the four configurations of potentials in Table \ref{tab:domain-limits}, one finds that the conditions upon the parameters are the same for configurations in the same diagonal. Therefore, the resulting conditions are the same whenever $\omS\muD$ has the same sign. Furthermore, if one considers all the particular cases contained in this description, two things can be noted: to obtain particular cases in which $\muD=0$, the limit from the right must be taken from configurations with $\muD>0$. To that it must be added the restriction that $\tilde{k}\gtrless\pm1$; as well, to obtain particular cases in which $\omS=0$, the limit must be taken from the left in the configurations in which $\omS<0$, added by the restriction $b\neq0$. Finally, to obtain the pure tensor-vector case ($\omS=\muD=0$), both limits must be taken, first for $\muD$ and then from $\omS$, from the configuration in which $\muD>0$ and $\omS<0$. Such particular cases will be explictly given in later sections of the text.
	
	To write the conditions, it is convenient to define the auxiliary quantities $\Lambda$ and $\Pi$, together with the functions $N(\tilde{k})$ and $M(\tilde{k})$:
	
	\begin{gather}
		\Lambda=\sqrt{\max(0,\tilde{\Lambda})} \;,\; \tilde{\Lambda}=1+2\muD+\dfrac{\muD}{\omS}b^2, 
	\\[5pt]
		\Pi=-\dfrac{1}{2b}\left[\dfrac{b^2}{\omS}\left(2+\dfrac{b^2}{\omS}\right) + \omS\muD \right] ,
	\\[5pt]
		N(\tilde{k})=-\frac{1}{2}+\dfrac{1}{4\sqrt{A\left(1+\dfrac{1-\tilde{k}^2}{\muD}\right)}}\left[\dfrac{1-\tilde{k}^2}{\muD}\left(2+\dfrac{1-\tilde{k}^2}{\muD}\right) +\omS\muD + 2\tilde{k}b \right] ,
	\\[5pt]
		M(\tilde{k})=-\frac{1}{2}+\dfrac{\sqrt{A\left(1+\dfrac{1-\tilde{k}^2}{\muD}\right)}}{2\omS\muD}\left[1-\tilde{k}^2 + \muD\left(1-\muD\sqrt{A\left(1+\dfrac{1-\tilde{k}^2}{\muD}\right)} \right)  \right].
	\end{gather}
	
	\noindent Then, the conditions are shown in Tables \ref{tab:roots+} and \ref{tab:roots-}.
	
	\begin{table}[h]
		\centering
		
		\begin{tabular}{c|c|c}
			&
			$-\Lambda \leq \tilde{k} \leq \Lambda$
			&
			$\tilde{k} \gtreqqless \pm \max\{\Lambda,\pm\Pi\}$
			\\
			\hline
			
			\makecell[c]{Subcriticality condition \\[-8pt] on singular potential}
			&
			$\begin{matrix*}
				\muD\geq-\dfrac{\omS}{2\omS+b^2}\text{, if }\left|b\right|<\sqrt{\left[\operatorname{sgn}(\omS)-1\right]\omS}
			\\[5pt]
				\forall \muD \text{, if } \left|b\right|\geq\sqrt{\left[\operatorname{sgn}(\omS)-1\right]\omS}
			\end{matrix*}$
			&
			$\forall (\muD,b)$
			\\
			\hline
			\hline
			\multicolumn{3}{c}{Number of real roots}
			\\
			\hline
			\hline
			
			No admissible domain
			& Not allowed
			& Not allowed
			\\
			\hline
			
			$0$
			& $\forall b$ , $0\leq n \leq \min\{N(\tilde{k}),M(\tilde{k})\}$
			& Not allowed
			\\
			\hline
			
			$1$
			& $\begin{matrix*}
				\forall b \;,\; N(\tilde{k})<n\leq M(\tilde{k})
			\\
				\forall b \;,\; n\geq\max\{N(\tilde{k}),M(\tilde{k})\} \;,\; n\neq M(\tilde{k})
			\end{matrix*}$ 
			& $b\lesseqqgtr 0$, $\forall n$
			\\
			\hline
			
			$2$
			& $\forall b$ , $M(\tilde{k})<n<N(\tilde{k})$
			& $b\gtreqqless0$, $\tilde{k}\neq\pm\Pi$, $\forall n$
			\\
			\hline
		\end{tabular}
		
		\caption{Conditions on $n$ and $\tilde{k}$ determining the number of real roots of the energy equation for $\omS\muD>0$. For the intervals of $\tilde{k}$ to be non-empty, the additional subcriticality conditions shown in the first row must be respected.}
		\label{tab:roots+}
	\end{table}


		\begin{table}[h]
		\centering
		
		\begin{tabular}{c|c|c}
			&
			$-\Lambda \leq \tilde{k} \leq \Lambda$
			&
			$\tilde{k} \gtreqqless \pm \max\{\Lambda,\pm\Pi\}$
			\\
			\hline
			
			\makecell[c]{Subcriticality condition \\[-8pt] on singular potential}
			&
			$\left|\muD\right|\leq\dfrac{\left|\omS\right|}{2\omS+b^2} \;,\; \forall b$
			&
			$\forall (\muD,b)$
			\\
			\hline
			\hline
			\multicolumn{3}{c}{Number of real roots}
			\\
			\hline
			\hline
			
			No admissible domain
			& $\forall (b,n)$
			& $\left\lbrace\begin{matrix*}[l]
				b\lesseqqgtr0 \;,\; \tilde{k}\neq\pm\Lambda \;,\; \forall n
			\\
			 (\tilde{k}^2-1-\muD)^2+2b\muD^2\tilde{k}\leq \muD^2(1-\omS\muD)
			\end{matrix*}\right.$
			\\
			\hline
			
			$0$
			& Not allowed
			& $\begin{matrix*}
				b\lesseqqgtr0 \;,\; n\geq N(\tilde{k})
			\\[5pt]
				b\gtreqqless0 \;,\; \tilde{k}\neq\pm\Pi \;,\; M(\tilde{k})\leq n \leq N(\tilde{k})
			\end{matrix*}$
			\\
			\hline
			
			$1$
			& Not allowed
			& $\begin{matrix*}
				b\lesseqqgtr 0 \;,\; n<\min\{N(\tilde{k},M(\tilde{k}))\}
			\\
				b\gtreqqless 0 \;,\; \tilde{k}\neq\pm\Pi \;,\; N(\tilde{k})\leq n < M(\tilde{k})
			\\
				b\gtreqqless 0 \;,\; \tilde{k}\neq\pm\Pi \;,\; n\geq\max\{N(\tilde{k}),M(\tilde{k})\} \;,\; n\neq N(\tilde{k})
			\end{matrix*}$
			\\
			\hline
			
			$2$
			& Not allowed
			& $b\gtreqqless0$, $n<\min\{N(\tilde{k}),M(\tilde{k})\}$
			\\
			\hline
		\end{tabular}
		
		\caption{Conditions on $n$ and $\tilde{k}$ determining the number of real roots of the energy equation for $\omS\muD<0$. For the intervals of $\tilde{k}$ to be non-empty, the additional subcriticality conditions shown in the first row must be respected.}
		\label{tab:roots-}
	\end{table}
	
		Therefore, the number of solutions for a given configuration of the potentials can be completely described in intervals of $\tilde{k}$ related to quantities that depend only on the potentials parameters, and intervals of $n$, which depends on functions of $\tilde{k}$ and the potentials parameters. Thus, the complete description of the spectrum of the generalized harmonic oscillator problem is achieved.
		
		Lastly, it remains to be determined wether a solution is related to a particle or an antiparticle state. To do so, it is instructive to introduce a common scale factor for all the potentials that connects the generalized problem to the free-particle case. Let it be  the scaling parameter $\lambda$, such that $0<\lambda\leq1$, which parametrizes the potentials parameters as follows: $b\rightarrow\lambda b$, $\omS\rightarrow\lambda^2\omS$ and $\muD\rightarrow\lambda\muD$. The energy equation function now is also a function of $\lambda$: $S=S(E,\lambda)$. Then, by taking the limit to the free case, $\lambda\rightarrow 0^{+}$, the roots of $S(E,0^+)$ are $\pm1$, the free particle and free antiparticle energies, respectively. Furthermore, for these roots, one has $S'(E,0^+)\gtrless0$ as well. Therefore, the sign of the first derivative in the root serves as a classification method for the particle and antiparticle sector in the free case. Since the scaling factor preserves continuously the ordering of the roots due to the convexity of the energy equation function inside $\mathcal{H}$, it follows that for the physical configuration, $\lambda=1$, this classification scheme still holds. Besides that, it can also be concluded that when there are both particle and antiparticle valid solutions, the particle one has energy greater or equal to the energy of the antiparticle state.
		
		As proven in Appendix \ref{ap:S(E)}, whenever $S(E)$ admits a stationary minimum, it is non-positive. From the convexity of the function, it follows then that if there are two real roots, this minimum is between them (regardless of it being part or not of the domain of the problem). Thus, it can be inferred, from Table \ref{TabRoots}, that the sign of $S'(L)$ and of $S'(R)$ must be equal to the signs of the left-most and right-most roots, respectively. Then, the classification scheme described above can be given equivalently in terms of the evaluation of derivatives of the boundaries of $\mathcal{H}$. Since the occurrence of less than two roots occur when the constraints exclude one or more solutions, this scheme also holds for when there is only one solution.
		
		 From that, it can be finally concluded, by inspection of Table \ref{TabRoots}, that for every configuration with $\omS\muD>0$ that has only one root, it is a particle solution if $\omS>0$ and an antiparticle solution when $\omS<0$. For the three conditions for there to be one root for $\omS\muD<0$, given in Table \ref{tab:roots-}, the first one is a particle solution and the two latter are antiparticle solutions when $\omS>0$, and the first is an antiparticle solution and the two latter are particle solutions when $\omS<0$. With that conclusion, the classification of all the solutions to the generalized harmonic oscillator problem is complete.

	\section{Particular cases}
	\label{sec:particular}
	
	In this section, the possible particular cases of the generalized problem solved above are considered. It will be shown that it recovers different particular cases already discussed in the literature. Furthermore, due to the completeness of the analysis made here for the solutions, the analysis for the particular cases can be developed as well, filling possible gaps in the literature.
	
	\subsection{Simple harmonic oscillator plus Dirac oscillator}
	
	One cannot simply take the singular potential parameter $\muD$ to be zero in the radial functions and energy equation to try to recover the full particular setup of a simple harmonic sum potential plus a Cornell tensor-vector potential. This would violate the construction of the \textit{Ansätze} (\ref{ansatzg}) and (\ref{ansatzf}), which required $\alpha$ and $\beta$ to be non-zero values, because the uncoupling condition (\ref{unccond}) would then imply that $\alpha$ is zero for $\bar{k}>-1/2$. Instead, on the radial function the particularization should be approached by taking a first order expansion of $\muD$ in $\delta$:
	
	\begin{gather}
		\delta=2\sqrt{\left(\bar{k}+\dfrac{1}{2}\right)^2 - \muD(1-E) }\approx \left|2\bar{k} +1\right| - \dfrac{\muD(1-E)}{\left|2\bar{k} +1\right|}.
	\end{gather}
	
	\noindent Using that expression to take the limit in the radial functions for $\bar{k}>-1/2$ results in
	
		\begin{gather}
		g(y)=2\bar{\alpha}\;y^{\frac{\bar{k}}{2}}e^{-y/2}L^{(\bar{k}-1/2)}_{n}(y),
		\\[5pt]
		f(y)=\frac{2\chi}{1+E}\bar{\alpha}\;y^{\frac{\bar{k}+1}{2}}e^{-y/2}\left[\left(\frac{a}{\chi^2}-1\right)L^{(\bar{k}-1/2)}_{n}(y) + 2L^{(\bar{k}+1/2)}_{n-1}(y) \right].
	\end{gather}
	
	\noindent For $\bar{k}<-1/2$ there is no special subtetly that requires attention.
	
	Regarding the spectrum analysis, for this particular case there is no exact general solution either, and the characteristics of the spectrum can be particularized from the general Tables \ref{tab:roots+} and \ref{tab:roots-}, taking the limit from the right of the configurations with $\muD>0$, and adding the restriction $\tilde{k}\gtrless\pm1$. For this limit and restriction, the functions $N(\tilde{k})$ and $M(\tilde{k})$ are not real anymore, therefore conditions for $n$ involving them are never satisfied and must be discarded. For $\muD=0$, $\Lambda=1$, thus the interval $-\Lambda \leq \tilde{k} \leq \Lambda$ is forbidden. The results are shown in Table \ref{tab:rootsu0}.
	
		\begin{table}[h]
		\centering
		
		\begin{tabular}{c|c}
			&
			$\tilde{k} \gtreqqless \pm \max\{1,\pm\Pi\}$
			\\
			\hline
			\hline
			\multicolumn{2}{c}{Number of real roots}
			\\
			\hline
			\hline
			
			No admissible domain
			& Not allowed
			\\
			\hline
			
			$0$
			& Not allowed
			\\
			\hline
			
			$1$
			& $b\lesseqqgtr 0$, $\tilde{k}\neq\pm1$ $\forall n$
			\\
			\hline
			
			$2$
			& $b\gtreqqless0$, $\tilde{k}\neq\pm\Pi$ , $\tilde{k}\neq\pm$1, $\forall n$
			\\
			\hline
		\end{tabular}
		
		\caption{Conditions on $n$ and $\tilde{k}$ determining the number of real roots of the energy equation for $\omS\neq0$ and $\muD=0$.}
		\label{tab:rootsu0}
	\end{table}
	
	From Table \ref{tab:rootsu0}, the conditions for the pure Dirac oscillator ($\omS=\muD=0$) can be recovered correctly as well. In this scenario, $\Pi\rightarrow\pm\infty$, for $b\lessgtr0$, and now $b\neq0$. Then, the condition for 1 root cannot be satisfied. Thus, as is known, the Dirac oscillator always binds both particles and antiparticles, such that for $b\gtrless0$ binds for $\tilde{k}\gtrless\pm1$, for any $n$.

	\subsection{The role of a singular potential in the Dirac oscillator problem}
	
	Taking now $\omS=0$ permits the investigation of the effects of the singular potential in the spectrum of the Dirac oscillator. This particular case is obtained by taking the limit from the left of the configuration in which $\omS<0$, and adding the restriction $b\neq0$. For this case, $\Lambda\rightarrow\pm\infty$ for $\muD\lessgtr0$, which forbids the either the interval $-\Lambda \leq \tilde{k} \leq \Lambda$ or $\tilde{k} \gtreqqless \pm \max\{\Lambda,\pm\Pi\}$, respectively. For the quantity $\Pi$, it tends to $\pm\infty$ for $b\lessgtr0$. For the function $M(\tilde{k})$, the limit results in
	
	\begin{gather}
		\lim_{\omS\rightarrow 0^-} M(\tilde{k})\rightarrow\left\lbrace\begin{matrix*}[l]
			\pm\infty \;,\; \tilde{k}^2\lessgtr 1+\muD\left(1-\dfrac{\muD\left|b\right|}{2} \right) 
		\\[5pt]
			-\dfrac{1}{2}-\dfrac{\muD}{4}\left(1+\dfrac{\muD\left|b\right|}{2} \right) \;,\; \tilde{k}^2=1+\muD\left(1-\dfrac{\muD\left|b\right|}{2} \right)
		\end{matrix*}\right. .
	\end{gather}
	
	Thus, the spectrum characteristics are shown in Tables \ref{tab:rootsw0-} and \ref{tab:rootsw0+}.
	
		\begin{table}[h]
		\centering
		
		\begin{tabular}{c|c}
			&
			$\forall\tilde{k}$
			\\
			\hline
			\hline
			\multicolumn{2}{c}{Number of real roots}
			\\
			\hline
			\hline
			
			No admissible domain
			& Not allowed
			\\
			\hline
			
			$0$
			& $b\neq0$ , $0\leq n \leq \min\{N(\tilde{k}),M(\tilde{k})\}$
			\\
			\hline
			
			$1$
			& $\begin{matrix*}
				b\neq0 \;,\; N(\tilde{k})<n\leq M(\tilde{k})
				\\
				b\neq0 \;,\; n\geq\max\{N(\tilde{k}),M(\tilde{k})\} \;,\; n\neq M(\tilde{k})
			\end{matrix*}$ 
			\\
			\hline
			
			$2$
			& $b\neq0$ , $M(\tilde{k})<n<N(\tilde{k})$
			\\
			\hline
		\end{tabular}
		
		\caption{Conditions on $n$ and $\tilde{k}$ determining the number of real roots of the energy equation for $\omS=0$ and $\muD<0$.}
		\label{tab:rootsw0-}
	\end{table}
	
			\begin{table}[h]
		\centering
		
		\begin{tabular}{c|c}
			&
			$\forall \tilde{k}$
			\\
			\hline
			\hline
			\multicolumn{2}{c}{Number of real roots}
			\\
			\hline
			\hline
			
			No admissible domain
			&  Not allowed
			\\
			\hline
			
			$0$
			& $b\neq0 \;,\; b\gtrless0 \;,\; \tilde{k}\neq\pm\Pi \;,\; M(\tilde{k})\leq n \leq N(\tilde{k})$
			\\
			\hline
			
			$1$
			& $\begin{matrix*}
				b\neq0 \;,\; b\gtrless 0 \;,\; \tilde{k}\neq\pm\Pi \;,\; N(\tilde{k})\leq n < M(\tilde{k})
				\\
				b\neq0 \;,\; b\gtrless 0 \;,\; \tilde{k}\neq\pm\Pi \;,\; n\geq\max\{N(\tilde{k}),M(\tilde{k})\} \;,\; n\neq N(\tilde{k})
			\end{matrix*}$
			\\
			\hline
			
			$2$
			& $b\neq0 \;,\; b\gtrless0$, $n<\min\{N(\tilde{k}),M(\tilde{k})\}$
			\\
			\hline
		\end{tabular}
		
		\caption{Conditions on $n$ and $\tilde{k}$ determining the number of real roots of the energy equation for $\omS=0$ and $\muD>0$.}
		\label{tab:rootsw0+}
	\end{table}
	
		For this configuration, there is no subcriticality condition anymore. For any intensity of the singular potential, there can always be a tensor-vector intensity strong enough to surpass the singular potential behaviour. 
	
	\section{Conclusion}
	\label{sec:conclusion}
	
	In this work, a complete computation of the analytical solutions and corresponding eigenergy equation for the bound states of a new generalized harmonic oscillator problem for a single spin-1/2 fermion involving the simultaneous presence of a simple harmonic oscillator, a singular potential, and a Cornell-type tensor potential (Coulomb-type plus Dirac oscillator) was achieved. By employing the same method of a previous paper, in which the tweaking of the multiplicative factor of the \textit{Ansätze} for the radial functions allow the exact decoupling of one of the radial equations, it was shown that they can be written in terms of generalized Laguerre polynomials, and a corresponding energy equation, which is irrational, can be found.
	
	Although the energy equation cannot lead generally to a direct expressions for the bound states spectrum (the corresponding algebraic equation if of eight order in the energy), methods of real analysis can be used to extract from the equation several informations about the behavior of the spectrum. By analyzing the boundaries of the domain in which the energy equation is physically relevant, conditions on the quantum numbers of the problem can be written in terms of the external potentials parameters, thus constituting a complete guide for the investigation of the solutions without knowing them beforehand. It was shown which conditions lead to the existence of none, one or two solutions for a given configuration of the potentials parameters. The critical conditions for the singular potential to allow the existence of bound-solutions was also discussed. Finally, a classification of wether a solution corresponds to a particle or antiparticle states was developed.
	
	The generalized problem solved in this paper emcompass some other problems already reported in the literature as particular cases. Not only the inspection of these particular cases sustain the validity of the general solutions presented here, but it is also an opportunity to fill some gaps in the literature regarding the analysis of the solutions and of the energy equation. Since the analysis presented here for the energy equation is more complete than most analysis presented in the literature, and are in a shape which can be directly particularized as well as the radial functions, the same analysis made for the generalized problem can be shown to these particular setups. Furthermore, the possibility of investigating the behaviour of a harmonic oscillator problem beyond the regime of the spin and pseudospin symmetries, which was allowed by the introduction of the singular potential is new, and widens the range of possible configurations for a spin-1/2 fermion system.
	
	\begin{acknowledgments}
		The study was financed in part by the Coordenação de Aperfeiçoamento de Pessoal de Nível Superior - Brasil (Capes) - Finance Code 001, Fundação de Amparo à Pesquisa do Estado de São Paulo (FAPESP) grant nº 2024/16575-7, and by FCT - Fundação para a Ciência e Tecnologia, I.P. in the framework of the projects UIDB/04564/2020 and UIDP/04564/2020, with DOI identifiers 10.54499/UIDB/04564/2020 and 10.54499/UIDP/04564/2020, respectively. PA would like to thank the São Paulo State University (Unesp), Guaratinguetá Campus, for supporting his stays at its Physics Department. AC and VM would like to thank Universidade de Coimbra for supporting their stays at its Physics Department.
	\end{acknowledgments}
 		
	\bibliographystyle{unsrt}
	\bibliography{/home/vbmendrot/Documentos/Zotero/refszotero.bib} 
	
	\appendix
	
	\section{Energy decomposition}
	\label{ap:decomp}
	
	One can project the Dirac spinor to select its upper or lower component with the projection operator $P^{\pm}=(1\pm\beta)/2$:
	
	\begin{gather}
		\Psi^+=P^+\Psi=\left( \begin{matrix}
			\varphi
			\\
			0
		\end{matrix}\right)
		\;,\;
		\Psi^-=P^-\Psi=\left( \begin{matrix}
			0\\
			\chi
		\end{matrix}\right),
	\end{gather}
	
	\noindent which can be used to rewrite the eigenvalue equation $H\Psi=\varepsilon\Psi$ --- in terms of the unscaled energy, $\varepsilon=mE$ ---, as two coupled equations
	
	\begin{gather}
		\bm{\alpha}\cdot(\bm{p} - \bm{A} - i\bm{U})\Psi_+=(\varepsilon + m - V_\Delta)\Psi_-,\label{psi+}\\
		\bm{\alpha}\cdot(\bm{p} - \bm{A} + i\bm{U})\Psi_-=(\varepsilon - m - V_\Sigma)\Psi_+.\label{psi-}
	\end{gather}
	
	\noindent From them, one can derive second order decoupled equations for each projection \cite{mendrotSymmetryGeneratorsQuantum2025}
	
	\begin{gather}\begin{aligned}\label{psi+2}
			\left\lbrace  \bm{p}^2 + (\Sigma_z A_\phi + U_{\tilde{\rho}})^2 - 2\dfrac{\Sigma_z A_\phi + U_{\tilde{\rho}}}{\tilde{\rho}}L_z\Sigma_z -\dfrac{1}{\tilde{\rho}} \dfrac{\partial}{\partial \tilde{\rho}}\left[\tilde{\rho}(\Sigma_z A_\phi + U_{\tilde{\rho}})\right] \right.& \\[10pt]
			\left.+ \dfrac{1}{\varepsilon + m - V_\Delta}\dfrac{\partial V_\Delta}{\partial \tilde{\rho}} \left[\dfrac{L_z\Sigma_z}{\tilde{\rho}} - \dfrac{\partial}{\partial \tilde{\rho}} - (\Sigma_z A_\phi + U_{\tilde{\rho}})\right]\right\rbrace  \Psi_+&\\[10pt]
			=\left( \varepsilon - m - V_\Sigma \right)\left( \varepsilon + m - V_\Delta \right) \Psi_+&,
		\end{aligned}
	\end{gather}
	
	\noindent and
	
	\begin{gather}\begin{aligned}\label{psi-2}
			\left\lbrace  \bm{p}^2 + (\Sigma_z A_\phi - U_{\tilde{\rho}})^2 + 2\dfrac{\Sigma_z A_\phi - U_{\tilde{\rho}}}{\tilde{\rho}}L_z\Sigma_z +\dfrac{1}{\tilde{\rho}} \dfrac{\partial}{\partial \tilde{\rho}}\left[\tilde{\rho}(\Sigma_z A_\phi - U_{\tilde{\rho}})\right] \right.& \\[10pt]
			\left.+ \dfrac{1}{\varepsilon - m - V_\Sigma}\dfrac{\partial V_\Sigma}{\partial \tilde{\rho}} \left[\dfrac{L_z\Sigma_z}{\tilde{\rho}} - \dfrac{\partial}{\partial \tilde{\rho}} - (\Sigma_z A_\phi - U_{\tilde{\rho}})\right]\right\rbrace  \Psi_-&\\[10pt]
			=\left( \varepsilon - m - V_\Sigma \right)\left( \varepsilon + m - V_\Delta \right) \Psi_-&.
		\end{aligned}
	\end{gather}
	
	From these equations we can determine the energy decomposition in terms of the expectation values equation
	
	\begin{gather}
		\langle \bm{p}^2\rangle
		+ \langle V_{A+U}\rangle
		+ \langle V_{\Delta\Sigma}\rangle
		+ \langle V_{\Delta AU}\rangle
		+ \langle V^{\Delta}_{\text{Darwin}}\rangle
		+ \langle V_{SO}\rangle= \varepsilon^2 - m^2,
		\label{exp+}
		\\[10pt]
		\langle \bm{p}^2\rangle
		+ \langle V_{A-U}\rangle
		+ \langle V_{\Delta\Sigma}\rangle
		+ \langle V_{\Sigma AU}\rangle
		+ \langle V^{\Sigma}_{\text{Darwin}}\rangle
		+ \langle V_{PSO}\rangle= \varepsilon^2 - m^2,
		\label{exp-}
	\end{gather}
	
	\noindent where
	
	\begin{gather}
		V_{A\pm U} =(\Sigma_zA_\phi \pm U_{\tilde{\rho}})^2
		- \frac{\Sigma_zA_\phi \pm U_{\tilde{\rho}}}{\tilde{\rho}}
		- \frac{d}{d\tilde{\rho}}(\Sigma_zA_\phi \pm U_{\tilde{\rho}}),
		\label{expAU}
		\\[5pt]
		V_{\Delta\Sigma}= V_\Delta V_\Sigma - \varepsilon(V_\Delta + V_\Sigma) + m(V_\Delta - V_\Sigma),
		\\[5pt]
		V_{\Delta AU}=-\dfrac{\Sigma_z A_\phi + U_{\tilde{\rho}}}{\varepsilon + m - V_\Delta}\dfrac{\partial V_\Delta}{\partial \tilde{\rho}}
		\quad,\quad
		V_{\Sigma AU}=-\dfrac{\Sigma_z A_\phi - U_{\tilde{\rho}}}{\varepsilon - m - V_\Sigma}\dfrac{\partial V_\Sigma}{\partial \tilde{\rho}},
		\\[5pt]
		V^{\Delta}_{\text{Darwin}}=-\dfrac{1}{\varepsilon + m - V_\Delta}\dfrac{\partial V_\Delta}{\partial \tilde{\rho}} \dfrac{\partial}{\partial \tilde{\rho}}
		\quad,\quad
		V^{\Sigma}_{\text{Darwin}}=-\dfrac{1}{\varepsilon - m - V_\Sigma}\dfrac{\partial V_\Sigma}{\partial \tilde{\rho}} \dfrac{\partial}{\partial \tilde{\rho}},
		\\[5pt]
		V_{SO}=\left[-2(\Sigma_zA_\phi + U_{\tilde{\rho}}) + \dfrac{1}{\varepsilon + m - V_\Delta}\dfrac{\partial V_\Delta}{\partial \tilde{\rho}}\right] \dfrac{L_z\Sigma_z}{\tilde{\rho}},
		\label{expSO}
		\\[5pt]
		V_{PSO}=\left[-2(\Sigma_zA_\phi - U_{\tilde{\rho}}) + \dfrac{1}{\varepsilon - m - V_\Sigma}\dfrac{\partial V_\Sigma}{\partial \tilde{\rho}}\right] \dfrac{L_z\Sigma_z}{\tilde{\rho}}.
		\label{expPSO}
	\end{gather}
	
	\noindent Throughout the calculations, it is required that these expectation values are finite, which shall lead to restrictions on the solutions parameters.
	
	\section{Classification of the number of roots of $S(E)$}
	\label{ap:S(E)}
	
	Inside the domain $\mathcal{D}$, the function
	
	\begin{gather}
		S(E)=E^2-\Gamma -\left(4n+2\sqrt{B(E)}\right) \sqrt{A(E)} \label{eqenergyap}
	\end{gather}
	
	\noindent has at most two real roots, which correspond to the eigenergies of the general harmonic oscillator problem. Since the sector of the full domain of the problem which can introduce a gap does so such that the entire gap is in the negative semi-plane of (\ref{eqenergyap}), it is irrelevant for the study of the roots. Furthermore, being (\ref{eqenergyap}) a convex function, the number of roots for a given configuration can be entirely determined by the sign of $S(E)$ and its derivative, $S'(E)$, in the boundaries of $\mathcal{H}$, given in (\ref{H}). These possible signs can be converted into conditions on the quantum numbers of the problem. The classification is given in Table \ref{TabRoots}, for $L<R$.
	
	\begin{table}[ht]
		\centering
		\begin{tabular}{c|c}
			\hline
			\textbf{Number of real roots } & \textbf{Signs of $S(E)$ and $S'(E)$ in the boundaries} \\
			\hline
			0 &
			\begin{tabular}{c}
				$S(L)\leq0$, $S(R)\leq0$, $\forall$ $S'(L)$, $S'(R)$ \\
				$S(L)\geq0$, $S(R)>0$, $S'(L)\geq0$, $S'(R)>0$ \\
				$S(L)>0$, $S(R)\geq0$, $S'(L)<0$, $S'(R)\leq0$
			\end{tabular}
			\\
			\hline
			1 &
			\begin{tabular}{c}
				$S(L)\leq0$, $S(R)>0$, $S'(L)<0$, $S'(R)>0$ \\
				$S(L)<0$, $S(R)>0$, $S'(L)\geq0$, $S'(R)>0$ \\
				$S(L)>0$, $S(R)\leq0$, $S'(L)<0$, $S'(R)>0$ \\
				$S(L)>0$, $S(R)<0$, $S'(L)<0$, $S'(R)\leq0$
			\end{tabular}
			\\
			\hline
			2 &
			$S(L)>0$, $S(R)>0$, $S'(L)<0$, $S'(R)>0$
			\\
			\hline
		\end{tabular}
		\caption{Combinations of signs of $S(E)$ and $S'(E)$ in the boundaries of $\mathcal{H}:(L,R)$, for $L<R$, that yields 0, 1 or 2 real roots of $S(E)$.}
		\label{TabRoots}
	\end{table}
	
	Besides that, there are also the possible cases for the domain $\mathcal{D}$ to be empty, which is another possibility of having no bound-states, but stemming from a different reason than there not being real roots inside the domain. These cases lead to a subcriticality condition for the singular potential parameter $\muD$. Whenever $L\geq R$, the domain is empty. For $L<R$, whenever there is a gap wider than a single point ($\Gamma>0$ for that to happen), such that both $L$ and $R$ are inside it, there is also an empty domain. This condition can be stated as: $L<R$, $L^2\leq\Gamma$ and $R^2\leq\Gamma$.
	
	\subsection{Impossible cases}
	
	One may argue that the condition for having 2 real roots could lead to there being only one root if the stationary minimum is exactly at zero. However, at such point, which may be called $E_0$, both $S(E_0)$ and $S'(E_0)$ are zero, then a Taylor expansion around such minimum has as first non-zero term $S''(E_0)(E-E_0)^2/2$. Therefore, this root has multiplicity two, and thus correpond to two solutions (particle and antiparticle) which have the same eingenergy.
	
	Another objection for 2 real roots condition is that it could lead to a 0 root case, if the stationary minimum is greater than zero. However, below there is a proof that whenever there is a stationary minimum of $S(E)$, it is necessarily non-positive.
	
	\textbf{Proposition:} $\forall E_0 \in \mathcal{H}|S'(E_0)=0 \implies S(E_0)\leq0$.
	
	\textbf{Proof:} If the proposition is proved for $n=0$, it must hold for any $n$. Let it be
	
	\begin{gather}
		F(E)=E^2-\Gamma -2\sqrt{A(E)B(E)},
	\end{gather}
	
	\noindent then $S(E)\leq F(E)$. Thus, it shall be proven that $F(E_0)\leq0$, and from it, immediately follows that the proposition is true.
	
	From (\ref{A}) and (\ref{B}), evaluated at the stationary minimum $E_0$, one can write
	
	\begin{gather}
		b^2\sqrt{\dfrac{B(E_0)}{A(E_0)}}=\sqrt{A(E_0)B(E_0)}-\omS\sqrt{\dfrac{B(E_0)}{A(E_0)}}(E_0+1)
	\\[5pt]
		\tilde{k}^2\sqrt{\dfrac{A(E_0)}{B(E_0)}}=\sqrt{A(E_0)B(E_0)}-\muD\sqrt{\dfrac{A(E_0)}{B(E_0)}}(E_0-1)
	\end{gather}
	
	\noindent Summing both equations and applying the inequality of the averages in the left-hand side, one gets
	
	\begin{gather}
		2\sqrt{A(E_0)B(E_0)} - \omS\sqrt{\dfrac{B(E_0)}{A(E_0)}}(E_0+1) - \muD\sqrt{\dfrac{A(E_0)}{B(E_0)}}(E_0-1) \geq 2b\tilde{k}.
	\end{gather}
	
	From $F'(E_0)=0$,
	
	\begin{gather}
	\muD\sqrt{\dfrac{A(E_0)}{B(E_0)}}=2E_0-\omS\sqrt{\dfrac{B(E_0)}{A(E_0)}}. \label{FE0}
	\end{gather}
	
	\noindent Substituing this in the inequality produces
	
	\begin{gather}
		2\sqrt{A(E_0)B(E_0)} - 2\omS\sqrt{\dfrac{B(E_0)}{A(E_0)}} - 2E_0(E_0-1) \geq 2b\tilde{k}.
	\end{gather}
	
	Now, adding the positive quantity $\left\lbrace E_0-1-\omS \left[B(E_0)/A(E_0)\right]^{1/2} \right\rbrace^2$ to the left side of the inequality results in
	
	\begin{gather}
		2\sqrt{A(E_0)B(E_0)} - 2E_0\omS\sqrt{\dfrac{B(E_0)}{A(E_0)}} - (E_0+1)(E_0-1) +\omS^2\dfrac{B(E_0)}{A(E_0)}\geq 2\tilde{k}b.
	\end{gather}
	
	Multiplying both sides of (\ref{FE0}) by $\omS\left[B(E_0)/A(E_0)\right]^{1/2}$ and using the resulting equation to substitute the last term in the left hand-side of the inequality, the desired result is achieved: $F(E_0)\leq 0$, and therefore the proposition is proved.
	
\end{document}